\documentclass[aps,prd,reprint,amsfonts,amssymb,amsmath,preprintnumbers,nofootinbib]{revtex4-1}
\usepackage[T1]{fontenc}
\usepackage[english]{babel}
\usepackage[Symbolsmallscale]{upgreek}
\usepackage{isomath}
\usepackage[protrusion=true,tracking=true,kerning=true,spacing=true,final,babel=true]{microtype}
\usepackage{physics}
\usepackage[final]{hyperref}

\begin{document}
\title{Can we detect quantum gravity with compact binary inspirals?}

\author{Alexander~C.~Jenkins}
\email{alexander.jenkins@kcl.ac.uk}
\affiliation{Theoretical Particle Physics and Cosmology Group, Physics Department, King's College London, University of London, Strand, London WC2R 2LS, United Kingdom}

\author{Andreas~G.~A.~Pithis}
\email{andreas.pithis@gmail.com}
\affiliation{Theoretical Particle Physics and Cosmology Group, Physics Department, King's College London, University of London, Strand, London WC2R 2LS, United Kingdom}

\author{Mairi~Sakellariadou}
\email{mairi.sakellariadou@kcl.ac.uk}
\affiliation{Theoretical Particle Physics and Cosmology Group, Physics Department, King's College London, University of London, Strand, London WC2R 2LS, United Kingdom}

\date{\today}
\preprint{KCL-PH-TH/2018-51}

\begin{abstract}
    Treating general relativity as an effective field theory, we compute the leading-order quantum corrections to the orbits and gravitational-wave emission of astrophysical compact binaries.
    These corrections are independent of the (unknown) nature of quantum gravity at high energies, and generate a phase shift and amplitude increase in the observed gravitational-wave signal.
    Unfortunately (but unsurprisingly), these corrections are undetectably small, even in the most optimistic observational scenarios.
\end{abstract}
\maketitle

\section{Introduction}
General relativity (GR) has passed an impressive range of observational tests in the weak-field (i.e., low-energy) regime~\cite{Will:2014kxa}.
However, it is well known that GR predicts the formation of singularities, indicating the breakdown of the theory in the extreme strong-field (high-energy) regime~\cite{Penrose:1964wq,Hawking:1966sx,Hawking:1969sw}.
Resolving such singularities is a major motivation for the construction of a theory of quantum gravity (QG)~\cite{DeWitt:1967yk}, which is well-behaved at high energies but is equivalent to GR in the low-energy limit.

The first attempts to quantise gravity failed due to the negative mass dimension of Newton's constant in 4D spacetime.
In the perturbative approach, this creates infinitely many divergences, whose renormalisation introduces infinitely many undetermined couplings, causing a loss of predictivity of the theory~\cite{tHooft:1974toh,Deser:1974zzd,Deser:1974zza,Goroff:1985th,vandeVen:1991gw}.

Despite this non-renormalisability, one can isolate the well-behaved low-energy regime of the theory from the divergences by integrating out the high-energy degrees of freedom~\cite{Donoghue:1994dn,Donoghue:1995cz,Burgess:2003jk}.
This results in an \emph{effective} field theory (EFT), which is a predictive, well-defined theory of QG at energies far below the Planck mass $M_\mathrm{P}$.
(This is analogous to the standard model of particle physics, which has been verified to exquisite precision up to $\sim10^{-16}M_\mathrm{P}$, but is expected to give way to new physics at higher energies.)

Remarkably, the dominant corrections to GR in this EFT are parameter-free, and therefore independent of the high-energy completion of QG.
One can thus calculate quantum corrections to low-energy gravitational phenomena, such as the Newtonian potential of two point masses~\cite{Donoghue:1993eb,BjerrumBohr:2002kt} and the classical Schwarzschild and Kerr spacetimes~\cite{BjerrumBohr:2002ks}.
These corrections are concrete, model-independent predictions of QG.
If observed, they would provide the first experimental evidence for the quantum nature of spacetime.\footnote{The observational signatures of the QG corrections to the two-body gravitational potential have already been investigated in the context of solar system dynamics in a series of papers by Battista et al, Refs.~\cite{Battista:2014oba,Battista:2014ija,Battista:2015zta,Battista:2015qxa,Battista:2015wwa}.}

In recent years, direct observations of gravitational waves (GWs) by the Advanced LIGO and Advanced Virgo interferometers~\cite{Abbott:2016blz,Abbott:2016nmj,Abbott:2017vtc,Abbott:2017gyy,Abbott:2017oio,TheLIGOScientific:2017qsa} have lead to powerful new tests of GR, including bounds on the GW propagation speed and searches for non-GR polarisation modes~\cite{TheLIGOScientific:2016src,Monitor:2017mdv,Callister:2017ocg,Abbott:2018utx}.
In light of these stringent new tests, it is pertinent to ask: \emph{Could deviations from GR due to quantum effects be observed with GW detectors?}
This Letter provides an answer, using the EFT of QG to compute the leading-order quantum corrections to compact binary (CB) inspirals, the most important class of source for current and future GW detectors.

\section{Quantum and relativistic corrections to the Newtonian potential}
In the EFT of QG, the one-loop scattering potential for two point masses $m_1,m_2$ in the harmonic gauge is~\cite{Donoghue:1993eb,BjerrumBohr:2002kt}
    \begin{equation}
    \label{eq:1-loop}
        V_\text{1-loop}\qty(r)=-\frac{GM^2\nu}{r}\qty(1+\frac{3}{2}\frac{r_\mathrm{S}}{r}+k\frac{\ell_\mathrm{P}^2}{r^2}),
    \end{equation}
    where $M\equiv m_1+m_2$ is the total mass, $\nu\equiv m_1m_2/M^2$ is the dimensionless reduced mass, $r_\mathrm{S}\equiv2GM/c^2$ is the Schwarzschild radius of the system, $\ell_\mathrm{P}\equiv\sqrt{\hbar G/c^3}$ is the Planck length, and $k=41/\qty(10\uppi)$ in the absence of other massless particles.\footnote{Note however that any additional massless particles will contribute to vacuum polarisation, altering the value of $k$ slightly. However, one still expects $k\sim1$.}
The three terms represent the classical Newtonian potential, the leading-order post-Newtonian (PN; i.e., relativistic) correction, and the leading-order QG correction, respectively.
PN corrections to CB inspirals have already been extensively studied to much higher order than shown here~\cite{Blanchet:2013haa}.
For this reason, we neglect all PN corrections and focus on the phenomenology of the quantum term, relative to the simple Newtonian case.
The point masses may represent extended objects here, as finite-size effects only appear at much higher order~\cite{Goldberger:2004jt}.

Even though the leading QG and PN terms both appear at the same loop order, the QG term is many orders of magnitude smaller.
This is because there are two fundamental length scales one can use to construct dimensionless terms in the $1/r$ expansion Eq.~\eqref{eq:1-loop}: the Planck length $\ell_\mathrm{P}$, and the Schwarzschild radius $r_\mathrm{S}$.
The two will coincide when $M\sim M_\mathrm{P}$, but we are interested in astrophysical objects with $M\gtrsim M_\odot\sim10^{38}M_\mathrm{P}$, so there exists a hierarchy of many orders of magnitude.
In fact, the PN and QG corrections in Eq.~\eqref{eq:1-loop} appear at the same loop order because they carry the same power of $G$, and therefore possess the same number of graviton vertices in the contributing Feynman diagrams.
From this viewpoint, it is clear that the two-loop potential will be of the form
    \begin{equation}
        V_\text{2-loop}\qty(r)=V_\text{1-loop}\qty(r)-\frac{GM^2\nu}{r}\qty(c_1\frac{r_\mathrm{S}^2}{r^2}+c_2\frac{r_\mathrm{S}\ell_\mathrm{P}^2}{r^3}+c_3\frac{\ell_\mathrm{P}^4}{r^4}),
    \end{equation}
    where the new terms have a factor of $G^2$ (hidden in the definitions of $r_\mathrm{S}$ and $\ell_\mathrm{P}$) compared to the tree-level Newtonian term.
Continuing in this way, one could in principle generate PN corrections to arbitrary order by extracting the $\qty(r_\mathrm{S}/r)^n$ terms.
These are relativistic effects, corresponding exactly to the PN corrections calculated with classical techniques.
The remaining terms contain powers of $\ell_\mathrm{P}$, representing QG effects.
In the regime $r\gg r_\mathrm{S}\gg\ell_\mathrm{P}$ these quantum corrections are dominated by the $\ell_\mathrm{P}^2/r^2$ term in Eq.~\eqref{eq:1-loop}.
We therefore study CB inspirals with the potential
    \begin{equation}
    \label{eq:qg-potential}
        V\qty(r)=-\frac{GM^2\nu}{r}\qty(1+k\frac{\ell_\mathrm{P}^2}{r^2}),
    \end{equation}
    comparing with the Newtonian potential to isolate the leading-order QG effects.

For reasons discussed above, there are many orders of PN corrections that are more significant than the leading-order QG corrections.
One can straightforwardly estimate which order PN is comparable to the leading-order QG term by setting $\ell_\mathrm{P}^2/r^2=\qty(r_\mathrm{S}/r)^n$ to give
    \begin{equation}
        n=2\frac{\ln\qty(\ell_\mathrm{P}/r)}{\ln\qty(r_\mathrm{S}/r)}.
    \end{equation}
For example, a binary with $m_1=m_2=M_\odot$ orbiting at $10~\mathrm{Hz}$ (i.e., as the binary is entering the LIGO-Virgo frequency band) would have leading-order QG effects equivalent to PN effects of order $n\approx44$.
(Cf. the current PN ``state of the art" of $n=4$~\cite{Blanchet:2013haa}.)
In the limit $r\to r_\mathrm{S}$, the PN corrections become large and $n$ diverges.
In the limit $r\to\infty$, the QG corrections are as large as the 2nd-order PN corrections, but both are negligible.

\section{Orbital perturbations}
In the absence of QG corrections we have the Keplerian two-body problem, in which the relative motion describes a constant elliptical orbit in a fixed plane, characterised by its semi-major axis (SMA) $a$ and eccentricity $e$.
The relative speed and separation of the bodies varies over each period $T$ of the orbit, so we define the mean motion $n\equiv2\uppi/T$ (i.e. the average angular velocity), which satisfies Kepler's equation,
    \begin{equation}
    \label{eq:Kepler}
        n=\sqrt{\frac{GM}{a^3}}.
    \end{equation}
The orientation of the orbital plane with respect to some fixed reference plane is determined by three angles: (i) the inclination $\iota$, which is the angle between the two planes; (ii) the argument of pericentre $\omega$, which is the angle in the orbital plane at which the binary reaches its minimum separation; (iii) the longitude of ascending node $\Omega$, which specifies the line where the two planes meet.
The quantities $a,e,n,\iota,\omega,\Omega$ are called the \emph{orbital elements}.

Once the QG corrections are taken into account, the resulting orbit is no longer a fixed ellipse.
We therefore define the ``osculating" orbital elements $a,e,n,\iota,\omega,\Omega$ as dynamical variables corresponding to the instantaneous ellipse defined by the relative speed and separation of the binary.
However, it is much simpler to calculate the corrected orbit in terms of some set of angle-action variables using Hamiltonian perturbation theory.
One such set is the Poincar\'e variables~\cite{2000ssd..book.....M}, with generalised co\"ordinates $\lambda,\gamma,z$ and conjugate momenta $\Lambda,\Gamma,Z$ defined by
    \begin{align}
    \label{eq:Poincare}
    \begin{split}
        \lambda&\equiv nt+\omega+\Omega,\\[5pt]
        \gamma&\equiv-\omega-\Omega,\\[5pt]
        z&\equiv-\Omega,
    \end{split}
    \begin{split}
        \Lambda&\equiv\sqrt{GM^3\nu^2a},\\
        \Gamma&\equiv\sqrt{GM^3\nu^2a}\qty(1-\sqrt{1-e^2}),\\
        Z&\equiv\sqrt{GM^3\nu^2a\qty(1-e^2)}\qty(1-\cos\iota).
    \end{split}
    \end{align}
The equations of motion (EoM) are then
    \begin{align}
    \label{eq:Hamilton}
    \begin{split}
        \dot{\lambda}&=\pdv{\mathcal{H}}{\Lambda},\\
        \dot{\Lambda}&=-\pdv{\mathcal{H}}{\lambda},
    \end{split}
    \begin{split}
        \dot{\gamma}&=\pdv{\mathcal{H}}{\Gamma},\\
        \dot{\Gamma}&=-\pdv{\mathcal{H}}{\gamma},
    \end{split}\qquad
    \begin{split}
        \dot{z}&=\pdv{\mathcal{H}}{Z},\\
        \dot{Z}&=-\pdv{\mathcal{H}}{z},
    \end{split}
    \end{align}
    where $\mathcal{H}$ is the Hamiltonian of the system.
The limit $e\to0$ corresponds to circular orbits, while the limit $\iota\to0$ corresponds to orbits in the reference plane; these must be taken after computing the EoM to produce the correct result.

In terms of the Poincar\'e variables, the Hamiltonian for the Newtonian case is simply
    \begin{equation}
        \mathcal{H}_\mathrm{N}=-\frac{G^2M^5\nu^3}{2\Lambda^2},
    \end{equation}
    so we find $\dot{\lambda}=G^2M^5\nu^3/\Lambda^3$.
Recalling the definitions in Eq.~\eqref{eq:Poincare}, we see that this reproduces Eq.~\eqref{eq:Kepler} with constant orbital elements, as expected.

We now include the leading-order QG correction to the Hamiltonian, read directly from Eq.~\eqref{eq:qg-potential},
    \begin{equation}
    \label{eq:Hamiltonian}
        \mathcal{H}=\mathcal{H}_\mathrm{N}+\mathcal{H}_\mathrm{QG},\qquad\mathcal{H}_\mathrm{QG}=-\frac{k\ell_\mathrm{P}^2GM^2\nu}{r^3},
    \end{equation}
The corrected Hamiltonian depends on the separation $r$, which has no closed-form expression in terms of the Poincar\'e variables, so the EoM cannot be derived through the straightforward application of Hamilton's equations.
One common solution is to time-average the Hamiltonian over an orbit, so that it describes the system's dynamics on timescales longer than $T$~\cite{2002clme.book.....G,1988macc.book.....A}.
The resulting low-frequency departures from Keplerian behaviour are called \emph{secular} perturbations to the system.
This is an acceptable approximation for small corrections to the Hamiltonian, as the resulting perturbations to the motion are usually negligible on timescales shorter than $T$.
It is sufficient to average over the Keplerian orbit rather than the corrected orbit, as the result is identical to leading order in the corrections.
The secular averaging operation is therefore
    \begin{equation}
    \label{eq:averaging}
        \ev{x}\equiv\int_{t_0}^{t_0+T}\frac{\dd{t}}{T}x=\int_{\psi_0}^{\psi_0+2\uppi}\frac{\dd{\psi}}{2\uppi}\frac{\qty(1-e^2)^{3/2}}{\qty(1+e\cos\psi)^2}x,
    \end{equation}
    where $\psi$ is the true anomaly (i.e., the angle of the orbit relative to $\omega$ in the orbital plane), and we have used the Keplerian equations for conservation of angular momentum and the orbital separation,
    \begin{equation}
    \label{eq:psi-dot-and-r}
        \dot{\psi}=n\sqrt{1-e^2}\qty(\frac{a}{r})^2,\qquad r=\frac{a\qty(1-e^2)}{1+e\cos\psi}.
    \end{equation}
Averaging the corrected Hamiltonian Eq.~\eqref{eq:Hamiltonian}, we find
    \begin{equation}
        \ev{\mathcal{H}}=\mathcal{H}_\mathrm{N}-\frac{k\ell_\mathrm{P}^2G^4M^{11}\nu^7}{\Lambda^3\qty(\Lambda-\Gamma)^3},
    \end{equation}
    so the secular EoM are
    \begin{align}
    \label{eq:sec-lambda-dot}
        \ev{\dot{\lambda}}&=\pdv{\ev{\mathcal{H}}}{\Lambda}=\frac{G^2M^5\nu^3}{\Lambda^3}+3k\ell_\mathrm{P}^2G^4M^{11}\nu^7\frac{2\Lambda-\Gamma}{\Lambda^4\qty(\Lambda-\Gamma)^4},\\
    \label{eq:sec-gamma-dot}
        \ev{\dot{\gamma}}&=\pdv{\ev{\mathcal{H}}}{\Gamma}=-\frac{3k\ell_\mathrm{P}^2G^4M^{11}\nu^7}{\Lambda^3\qty(\Lambda-\Gamma)^4}.
    \end{align}
Since $\pdv*{\ev{\mathcal{H}}}{z}=0$, we see from Eq.~\eqref{eq:Hamilton} that $Z$ is constant on long timescales; it may undergo oscillations during each orbit, but these vanish when performing the secular averaging.
Using Eq.~\eqref{eq:Poincare}, we therefore set $\Omega=\iota=0$ and fix the orbit within the reference plane without loss of generality.
The other momenta $\Lambda,\Gamma$ are also conserved in the secular Hamiltonian, so that $a$ and $e$ are constant on long timescales.
Equations~\eqref{eq:Poincare} and~\eqref{eq:averaging} give
    \begin{equation}
        \ev{\dot{\lambda}+\dot{\gamma}}=\ev{\dv{\qty(nt)}{t}}=\left.n\right|_{t_0+T}+\frac{t_0}{T}\qty(\left.n\right|_{t_0+T}-\left.n\right|_{t_0}),
    \end{equation}
    but the lhs is constant and independent of the arbitrary choice of $t_0$, which implies the same for the rhs, so $n$ is constant.
(This is also true on short timescales, unlike the conservation of $a$, $e$, and $\iota$.)
Summing Eqs.~\eqref{eq:sec-lambda-dot} and~\eqref{eq:sec-gamma-dot} and rewriting in terms of the orbital elements, we therefore have
    \begin{equation}
    \label{eq:sec-n}
        n=\sqrt{\frac{GM}{a^3}}\qty[1+\frac{3k\ell_\mathrm{P}^2}{a^2\qty(1-e^2)^{3/2}}].
    \end{equation}
Comparing with Eq.~\eqref{eq:Kepler}, we see that the QG correction causes the binary to orbit slightly faster.
Intuitively, this is necessary to counteract the slightly stronger gravitational attraction between the bodies.
Similarly, rewriting Eq.~\eqref{eq:sec-gamma-dot} in terms of orbital elements gives
    \begin{equation}
    \label{eq:sec-omega-dot}
        \ev{\dot{\omega}}=\sqrt{\frac{GM}{a^3}}\frac{3k\ell_\mathrm{P}^2}{a^2\qty(1-e^2)^2},
    \end{equation}
    so the faster mean motion causes the pericentre to advance at an average rate Eq.~\eqref{eq:sec-omega-dot}, many orders of magnitude smaller than the corresponding relativistic precession.
For example, Eq.~\eqref{eq:sec-omega-dot} predicts a perihelion advance of $\sim10^{-84}$ arcseconds per century for the Mercury-Sun system, cf. 43 arcseconds per century due to relativistic effects~\cite{Will:2014kxa}.

\section{Orbital decay through GW emission}
We now calculate the GW emission from the QG-corrected CB orbit, using the quadrupole formula.
Adopting Cartesian co\"ordinates $x_i$ in the centre-of-mass frame, with the orbit in the $x_1$-$x_2$ plane, the rate of energy loss is~\cite{Maggiore:1900zz}
    \begin{equation}
    \label{eq:quadrupole-formula}
        \dot{E}=-\frac{2G}{15c^5}\qty(\dddot{\mathcal{M}}^2_{11}+\dddot{\mathcal{M}}^2_{22}+3\dddot{\mathcal{M}}^2_{12}-\dddot{\mathcal{M}}_{11}\dddot{\mathcal{M}}_{22}),
    \end{equation}
    where $\mathcal{M}_{ij}=M\nu x_ix_j$ is the second mass moment (under the point-mass approximation).
For our corrected orbit, this becomes
    \begin{align}
    \begin{split}
    \label{eq:second-mass-moments}
        \mathcal{M}_{11}&=M\nu r^2\cos^2(\psi+\omega),\\
        \mathcal{M}_{12}=\mathcal{M}_{21}&=M\nu r^2\cos(\psi+\omega)\sin(\psi+\omega),\\
        \mathcal{M}_{22}&=M\nu r^2\sin^2(\psi+\omega),
    \end{split}
    \end{align}
    where Eq.~\eqref{eq:psi-dot-and-r} holds as before, but with the corrected value of $n$ from Eq.~\eqref{eq:sec-n}, and $\omega$ evolves according to Eq.~\eqref{eq:sec-omega-dot}.
To give a gauge-invariant notion of GW energy, we must average $\dot{E}$ over one (QG-corrected) orbit.
While GW emission will cause $a,e,n,\ev{\dot{\omega}}$ to evolve, we assume that the energy radiated on orbital timescales $T$ is much less than the energy of the orbit so that
    \begin{equation}
        \frac{T\dot{n}}{n}\approx\frac{T\ddot{\omega}}{\dot{\omega}}\approx\frac{T\dot{a}}{a}\approx\frac{T\dot{e}}{e}\approx0.
    \end{equation}
Applying Eqs.~\eqref{eq:quadrupole-formula} and~\eqref{eq:second-mass-moments} with $a,e,n,\ev{\dot{\omega}}$ constant, and inserting the QG-corrected values of $n$ and $\ev{\dot{\omega}}$ from Eqs.~\eqref{eq:sec-n} and~\eqref{eq:sec-omega-dot}, we find
    \begin{align}
    \begin{split}
    \label{eq:energyloss}
        \ev{\dot{E}}=-&\frac{32G^4M^5\nu^2}{5c^5a^5\qty(1-e^2)^{7/2}}\qty(1+\frac{73}{24}e^2+\frac{37}{96}e^4)\\
        &\times\qty[1+\frac{30k\ell_\mathrm{P}^2}{a^2\qty(1-e^2)^{3/2}}\qty(\frac{1+\frac{37}{20}e^2-\frac{31}{160}e^4}{1+\frac{73}{24}e^2+\frac{37}{96}e^4})].
    \end{split}
    \end{align}
As expected, we recover the classical Peters-Mathews formula in the limit $\ell_\mathrm{P}\to0$~\cite{Peters:1963ux}.
The $\order{\ell_\mathrm{P}^2/a^2}$ term represents additional GW power due to the strengthening of the attractive force.
Similarly, using the equation for angular momentum loss under GW emission~\cite{Maggiore:1900zz},
    \begin{equation}
        \dot{L}=-\frac{2G}{5c^5}\qty[\ddot{\mathcal{M}}_{12}\qty(\dddot{\mathcal{M}}_{11}-\dddot{\mathcal{M}}_{22})-\dddot{\mathcal{M}}_{12}\qty(\ddot{\mathcal{M}}_{11}-\ddot{\mathcal{M}}_{22})],
    \end{equation}
    we find
    \begin{align}
    \begin{split}
    \label{eq:angmomloss}
        \ev{\dot{L}}=-&\frac{32G^{7/2}M^{9/2}\nu^2}{5c^5a^{7/2}\qty(1-e^2)^2}\qty(1+\frac{7}{8}e^2)\\
        &\times\qty[1-\frac{27k\ell_\mathrm{P}^2}{a^2\qty(1-e^2)^{3/2}}\qty(\frac{1+\frac{7}{72}e^2-\frac{1}{18}e^4}{1+\frac{7}{8}e^2})].
    \end{split}
    \end{align}

To translate Eqs.~\eqref{eq:energyloss} and~\eqref{eq:angmomloss} into expressions for $\dot{a},\dot{e}$, we write $E,L$ in terms of orbital elements, accounting for $\order{\ell_\mathrm{P}^2/a^2}$ corrections.
By definition we have
    \begin{equation}
        L\equiv M\nu\vb*r\cross\dot{\vb*r}=M\nu r^2\qty(\dot{\psi}+\dot{\omega}).
    \end{equation}
Differentiating gives $\dot{L}=M\nu\vb*r\cross\ddot{\vb*r}$, which vanishes identically when GW emission is neglected, as the gravitational acceleration $\ddot{\vb*r}$ is parallel to the separation vector $\vb*r$.
However, back-reaction due to GW emission causes a small acceleration perpendicular to $\vb*r$, giving $\dot{L}<0$ as we found above.
For all but the tightest orbits, $|L/\dot{L}|\gg T$, so we can safely treat $L$ as constant when performing the secular averaging.
We therefore find
    \begin{equation}
    \label{eq:angmom}
        L=\sqrt{GM^3\nu^2a\qty(1-e^2)}\qty[1+3k\frac{\ell_\mathrm{P}^2}{a^2}\frac{1+\sqrt{1-e^2}}{\qty(1-e^2)^2}],
    \end{equation}
    which is greater than the Keplerian value, due to the faster mean motion and the pericentre advance.
Now we write
    \begin{align}
    \begin{split}
    \label{eq:energy-def}
        E&\equiv\frac{1}{2}M\nu\qty|\dot{\vb*r}|^2+V\qty(r)\\
        &=\frac{1}{2}M\nu\dot{r}^2+\frac{L^2}{2M\nu r^2}-\frac{GM^2\nu}{r}\qty(1+k\frac{\ell_\mathrm{P}^2}{r^2}),
    \end{split}
    \end{align}
    where Eq.~\eqref{eq:psi-dot-and-r} gives $\dot{r}=aen\sin\psi/\sqrt{1-e^2}$.
Averaging Eq.~\eqref{eq:energy-def} over one orbit, and substituting the corrected values for $n$ and $L$, we obtain
    \begin{equation}
    \label{eq:energy}
        E=-\frac{GM^2\nu}{2a}\qty[1-\frac{10k\ell_\mathrm{P}^2}{a^2\qty(1-e^2)^{3/2}}],
    \end{equation}
    which is slightly larger than the Keplerian value, making it slightly easier to gravitationally unbind the CB than in the Newtonian case.

Combining Eqs.~\eqref{eq:angmom} and~\eqref{eq:energy} with Eqs.~\eqref{eq:energyloss} and~\eqref{eq:angmomloss}, we find the secular evolution of the SMA and eccentricity,
    \begin{align}
    \begin{split}
    \label{eq:adot}
        \ev{\dot{a}}=-&\frac{64G^3M^3\nu}{5c^5a^3\qty(1-e^2)^{7/2}}\qty(1+\frac{73}{24}e^2+\frac{37}{96}e^4)\\
        &\times\qty[1+\frac{60k\ell_\mathrm{Pl}^2}{a^2\qty(1-e^2)^{3/2}}\frac{1+\frac{397}{240}e^2-\frac{421}{1920}e^4}{1+\frac{73}{24}e^2+\frac{37}{96}e^4}],
    \end{split}\\
    \begin{split}
    \label{eq:edot}
        \ev{\dot{e}}=-&\frac{304G^3M^3\nu e}{15c^5a^4\qty(1-e^2)^{5/2}}\qty(1+\frac{121}{304}e^2)\\
        &\times\Bigg\{1+\frac{468k\ell_\mathrm{P}^2}{19a^2e^2\qty(1-e^2)^{3/2}\qty(1+\frac{121}{304}e^2)}\bigg[1+\frac{133}{156}e^2\\
        &\quad-\frac{211}{1248}e^4+\frac{1}{52}e^6-\frac{3}{26}\sqrt{1-e^2}\qty(1+\frac{7}{8}e^2)\bigg]\Bigg\},
    \end{split}
    \end{align}
    which match the classical results when $\ell_\mathrm{P}\to0$~\cite{Maggiore:1900zz}.
Equations~\eqref{eq:adot} and~\eqref{eq:edot} form a coupled system that is analytically intractable due to its non-linearity, and numerically intractable due to the vastly fundamental different time- and length-scales, which require extreme numerical precision to resolve.
However, we can make progress by looking at circular orbits, $e\to0$.
Taking $e\ll1$ is well justified, as the majority of CBs in the LIGO-Virgo frequency band are thought to form through common evolution, with low eccentricity as a result.
However, neglecting eccentricity in Eqs.~\eqref{eq:adot} and~\eqref{eq:edot} requires $e\ll\ell_\mathrm{P}/a$, which is much more restrictive than $e\ll1$, and is almost certainly false for realistic binaries.
Nonetheless, for an order-of-magnitude estimate of the QG corrections one can neglect eccentricity entirely.
We therefore take $e\to0$ in Eq.~\eqref{eq:adot} to obtain
    \begin{equation}
    \label{eq:adot-circular}
        \dot{a}=-\frac{64\nu}{5c^5}\qty(\frac{GM}{a})^3\qty(1+60k\frac{\ell_\mathrm{P}^2}{a^2}).
    \end{equation}
We have dropped the angle brackets, since the orbital separation remains fixed in the $e\to0$ case, and there is no further need to perform the secular averaging.

In the $e\to0$ case, the precession of the circular orbit due to $\dot{\omega}$ is indistinguishable from the mean motion of the bodies around that orbit, so that $\dot{\omega}$ is absorbed into $n$.
Combining Eqs.~\eqref{eq:sec-n} and~\eqref{eq:sec-omega-dot}, we obtain
    \begin{equation}
    \label{eq:n-circular}
        \left.n\right|_{e=0}=\sqrt{\frac{GM}{a^3}}\qty(1+6k\frac{\ell_\mathrm{P}^2}{a^2})
    \end{equation}

\section{Gravitational waveform and observational prospects}
Using the quadrupole formula, we write the time-domain complex GW waveform observed at a distance $R$ along the orbital axis (i.e. viewing the binary face-on) as~\cite{Maggiore:1900zz}
    \begin{equation}
        h\qty(t)\equiv h_+\qty(t)-\mathrm{i}h_\times\qty(t)=\frac{G}{c^4R}\qty(\ddot{\mathcal{M}}_{11}-2\mathrm{i}\ddot{\mathcal{M}}_{12}-\ddot{\mathcal{M}}_{22}).
    \end{equation}
Using Eqs.~\eqref{eq:second-mass-moments},~\eqref{eq:adot-circular}, and~\eqref{eq:n-circular}, this becomes
    \begin{equation}
        h\qty(t)=\mathcal{A}\qty(t)\exp[\mathrm{i}\Psi\qty(t)],
    \end{equation}
    which we have written in terms of a phase,
    \begin{equation}
        \Psi\qty(t)\equiv\Psi_0+2\int_{t_0}^t\dd{t'}n\qty(t'),
    \end{equation}
    and an amplitude,
    \begin{equation}
        \mathcal{A}\qty(t)=\frac{4\nu\qty(GM)^{5/3}n^{2/3}\qty(t)}{c^4R}\qty[1+\frac{6k\ell_\mathrm{P}^2n^{4/3}\qty(t)}{\qty(GM)^{2/3}}].
    \end{equation}
This defines two types of QG correction to the waveform: (i) a phase shift, and (ii) an amplitude increase.
We calculate the phase shift by writing
    \begin{equation}
        \int_{t_0}^t\dd{t'}n\qty(t')=\int_{a_0}^{a\qty(t)}\dd{a}\frac{n}{\dot{a}},
    \end{equation}
    and substituting the expressions in Eqs.~\eqref{eq:adot-circular} and~\eqref{eq:n-circular} to give
    \begin{align}
    \begin{split}
        \Psi\qty(t)=\Psi_0&+\frac{1}{16\nu}\qty[\qty(\frac{\uppi r_\mathrm{S}f_0}{2c})^{-5/3}-\qty(\frac{\uppi r_\mathrm{S}f\qty(t)}{2c})^{-5/3}]\\
        &-\frac{65k\ell_\mathrm{P}^2}{\nu r_\mathrm{S}^2}\qty[\qty(\frac{\uppi r_\mathrm{S}f_0}{2c})^{-1/3}-\qty(\frac{\uppi r_\mathrm{S}f\qty(t)}{2c})^{-1/3}],
    \end{split}
    \end{align}
    where $f\qty(t)\equiv n\qty(t)/\uppi$ is the GW frequency.
The classical expression is regained by taking $\ell_\mathrm{P}\to0$, so the phase shift is
    \begin{equation}
        \updelta\Psi_\mathrm{QG}\qty(t)\equiv\Psi\qty(t)-\lim_{\ell_\mathrm{P}\to0}\Psi\qty(t),
    \end{equation}
    where the comparison is between the Newtonian and QG-corrected waveforms at the same initial and final frequencies.
If a binary is observed for long enough then $f\qty(t)\gg f_0$ and the phase shift is simply
    \begin{equation}
    \label{eq:phase-shift}
        \updelta\Psi_\mathrm{QG}=-\frac{65k\ell_\mathrm{P}^2}{\nu r_\mathrm{S}^2}\qty(\frac{\uppi r_\mathrm{S}f_0}{2c})^{-1/3},
    \end{equation}
    which is greater for low-mass binaries and for lower initial frequencies, due to the longer signal duration.
Similarly, we find an amplitude increase
    \begin{equation}
    \label{eq:amplitude-shift}
        \updelta\mathcal{A}_\mathrm{QG}\qty(t)=\frac{12\uppi^2\nu k\ell_\mathrm{P}^2r_\mathrm{S}}{c^2R}f^2\qty(t),
    \end{equation}
    which grows as the binary approaches merger.

Of the two effects calculated above, the phase shift is more easily observable, for two reasons.
Firstly, even if the source's host galaxy is identified, small corrections to the amplitude as in Eq.~\eqref{eq:amplitude-shift} are dwarfed by the statistical uncertainty of $R$.
Secondly, matched-filter searches using waveform templates are very sensitive to the signal's phase evolution, as any loss of phase coherence between signal and template causes destructive interference and a loss of statistical significance in the search.
The precision with which one can measure the phase is inverse to the signal-to-noise ratio (SNR)~\cite{Creighton:2011zz}, so a very loud signal of, say, $\mathrm{SNR}=100$ would allow a phase precision of $\sim10^{-2}$ radians.\footnote{Cf. the first detected signal, GW150914, which had $\mathrm{SNR}\approx24$.}
An ensemble of $N$ measurements would enhance this by a factor of $1/\sqrt{N}$, so that using 100 such signals one could detect a phase shift of $\sim10^{-3}$ radians.

Even in this highly optimistic scenario, it is clear from Eq.~\eqref{eq:phase-shift} that the effect is many orders of magnitude too small to be detected for any astrophysical signal.
E.g., a binary with $m_1=m_2=M_\odot$ observed from an initial frequency of $f_0=10~\mathrm{Hz}$ until merger (i.e., the full frequency window of ground-based interferometers) would undergo a phase shift of just $\sim10^{-74}$ radians.
Any system with mass low enough to produce a measureable phase shift would produce a signal far too weak to be detected in the first place.

\section{Discussion}
Equation~\eqref{eq:adot} shows that the QG corrections will likely be larger for eccentric CBs, $e>0$, which we have not considered here.
However, given the gap of more than 70 orders of magnitude between the QG corrections for the $e=0$ case and the optimistic phase sensitivity, it seems very unlikely that this will lead to a detectable effect for any reasonable value of the eccentricity.

We have focused on the modifications to the GW signal due to the perturbed dynamics of the binary, and have neglected any modifications to the quadrupole formula [Eq.~\eqref{eq:quadrupole-formula}] or the GW propagation.
It is also possible to calculate these other modifications within the same EFT approach we have adopted, as shown recently in Refs.~\cite{Calmet:2018qwg,Calmet:2018rkj}.
These papers found that in addition to the classical massless graviton, the EFT of QG predicts two massive propagating modes, which have an alternative dispersion relation, and whose production is described by an additional term in the quadrupole formula.
However, experimental bounds on the masses of these additional modes imply that they can only be radiated from a CB with an orbital frequency greater than $\approx10^{13}~\mathrm{Hz}$, which is unattainable for any astrophysical CB.
For orbital frequencies less than this, the GW emission will be purely given by the classical graviton, with the classical quadrupole formula as in Eq.~\eqref{eq:quadrupole-formula}.
Thus for any astrophysical source, the leading corrections to the GW signal from the EFT of QG will be the ones presented here.

Note however that there may be non-perturbative QG effects which cannot be predicted within EFT, but may have observational consequences.
In particular, there has been much recent interest in the proposal of `BH echoes' in the CB ringdown signal due to quantum modifications to the structure of the BH event horizon~\cite{Cardoso:2016oxy,Abedi:2016hgu,Tsang:2018uie}.
These effects, however, are very speculative and model-dependent.
Our results show that QG is \emph{generally} unobservable with CB inspirals, but there may still be particular high-energy completions of QG that give rise to observable non-perturbative phenomena, such as echoes.

\section{Conclusion}
We have used results from the EFT of QG~\cite{Donoghue:1994dn,Donoghue:1995cz,Burgess:2003jk,Donoghue:1993eb,BjerrumBohr:2002kt} to compute leading-order quantum corrections to the orbits and GW emission of CBs.
By virtue of the EFT approach, these corrections are independent of the (unknown) nature of QG at high energies---if gravity is indeed quantised, then the dominant quantum effects for CBs will be of the form presented here.
The QG correction to the Newtonian potential modifies the inspiral orbit, leading to a phase shift and amplitude increase in the observed GW signal, given by Eqs.~\eqref{eq:phase-shift} and~\eqref{eq:amplitude-shift}.
Unfortunately (but unsurprisingly) these corrections are undetectably small, even in the most optimistic observational scenarios.

\begin{acknowledgments}
A.C.J. and A.G.A.P. are supported by King's College London through Graduate Teaching Scholarships.
M.S. is supported in part by the Science and Technology Facility Council (STFC), United Kingdom, under the research grant ST/P000258/1.
\end{acknowledgments}

\bibliography{qg-gw}
\end{document}